\documentclass{cargese}\usepackage{graphicx}
\let\footnote\savefootnote
\let\footnotetext\savefootnotetext 
\let\footnoterule\savefootnoterule 
\normallatexbib

\def\beq{\begin{equation}}
\def\eeq{\end{equation}}

\begin{document}
\articletitle[Non-BPS D-branes]{Anomalous couplings of non-BPS \\ D-branes}
\author{Marco Bill\'o}
\affil{Dipartimento di Fisica Teorica, Universit\`a di
Torino and\\
I.N.F.N., Sezione di Torino, via P. Giuria 1, I-10125, Torino, Italy }         
\author{Ben Craps and Frederik Roose }
\affil{Instituut voor Theoretische Fysica\\
Katholieke Universiteit Leuven, B-3001 Leuven, Belgium }
\begin{abstract}
Non-BPS type II D-branes couple to R-R potentials via an action that,  
upon tachyon condensation, gives rise to the Wess-Zumino action of BPS D-branes.
\end{abstract}
\index{non-BPS D-branes}
Non-BPS branes in string theory have received a lot of attention after the seminal
work by Sen \cite{sen}, as such non-BPS states allow for highly non-trivial
tests of string dualities. In a related development, unstable and/or non-BPS
D-branes have played a key role in the unifying description of all 
lower-dimensional D-branes as non-trivial excitations on suitable 
configurations of higher-dimensional branes.
This description exhibits in a constructive way the fact that D-brane charges 
take values in appropriate K-theory groups of space-time.
\par
For type IIB this was demonstrated by Witten in \cite{witten}, where 
all branes were built from sufficiently many D9--anti-D9
pairs. In this set-up, the D$p$-brane that emerges via a process of
tachyon condensation inherits its anomalous Wess-Zumino couplings to the R-R 
fields \index{Wess-Zumino action} \cite{GHM}, 
\beq \label{wzbps}
S_{\rm WZ}=\frac{T_p}{\kappa}\int_{p+1} C\wedge {\rm Tr}\, 
{\rm e}^{2\pi\alpha'\,F+B}\wedge
\sqrt{\hat{A}(R_T)/\hat{A}(R_N)}~,
\eeq
from the analogous couplings of the parent branes. In eq.~(\ref{wzbps}),
$T_p/\kappa$ denotes the D$p$-brane tension, $C$ a formal sum of R-R 
potentials, $F$ the gauge field on  the brane and $B$ the NS-NS two-form.
The trace is over the Chan-Paton indices.
Further, $R_T$ and $R_N$ are the curvatures of the tangent and normal 
bundles of the D-brane world-volume, and $\hat{A}$ denotes the A-roof genus.
\par
As for type IIA, Horava described in \cite{horava} 
how to construct BPS D($p-2k-1$)-branes as 
bound states of (sufficiently many) unstable D$p$-branes (and thus
how to construct all type IIA D-branes in terms of D9-branes). 
The lower-dimensional BPS
branes arise as the result of the condensation of a tachyon field 
into a vortex configuration, accompanied by non-trivial gauge fields.
In this case, it was not clear how the lower-dimensional BPS branes 
acquire the R-R couplings of eq.~(\ref{wzbps}).
\par
In \cite{ournonbps} we argued that all type II non-BPS branes couple 
universally to Ramond-Ramond fields as given by\footnote{%
One term of this action (the one describing the coupling of a non-BPS D9-brane 
to $C_9$) was conjectured to be present in Ref.~\cite{horava}
and obtained by a disc computation 
in Ref.~\cite{sen}.}
\index{Wess-Zumino action}
\beq \label{wznonbps}
S'_{\rm WZ}=a\int_{p+1} C\wedge d\,{\rm Tr}\, T \, 
{\rm e}^{2\pi\alpha'\,F+B}\wedge
\sqrt{\hat{A}(R_T)/\hat{A}(R_N)}~,
\eeq
where $T$ is the real, adjoint tachyon field living on the non-BPS brane and 
$a$ is a constant.
Let us indicate how the R-R couplings of eq.~(\ref{wznonbps}) account 
for the R-R couplings (\ref{wzbps}) of the stable lower-dimensional brane 
that emerges from the process of tachyon condensation.
The cases D$p$ $\rightarrow$ D($p-1$) 
and D$p$ $\rightarrow$ D($p-3$) will be treated in detail. It will turn out, for instance,
that the R-R charges of the D8-branes and D6-branes one constructs from unstable D9-branes 
\cite{horava} have the expected ratio.
\par
Consider first a single non-BPS D$p$-brane. There is a  real
tachyon field living on its worldvolume. The tachyon potential 
is assumed to be such that the vacuum manifold consists of the two points 
$\{T_0,-T_0\}$. The tachyon 
can condense in a non-trivial (anti)-kink configuration $T(x)$ depending on a
single coordinate. 
The R-R coupling (\ref{wznonbps}) on the D$p$-brane reads in this case 
\begin{equation}
\label{m1}
a\int_{p+1} C\wedge dT\wedge {\rm e}^{2\pi\alpha' F + B} \wedge 
\sqrt{\hat{A}(R_T)/\hat{A}(R_N)}~.
\end{equation}
It involves the topological density  $\partial_x T(x)$, which is localized at 
the core of the kink and is such that $\int dT(x) = \pm 2 T_0$. 
In the limit of zero size, $dT(x) = 2 T_0\delta(x-x_0) dx$, and the above 
action takes the form of the usual Wess-Zumino
effective action for a BPS D($p-1$)-brane, localized in the $x$-direction at
$x_0$:
\begin{equation}
\label{m2}
2T_0a\int_{p} C\wedge {\rm e}^{2\pi\alpha' F + B} 
\wedge \sqrt{\hat{A}(R_T)/\hat{A}(R_N)}~.
\end{equation}
\par
As a less trivial example, 
let us start from two unstable D$p$-branes. The tachyon field $T$,
transforming in the adjoint of the ${\rm U}(2)$ gauge group, 
can form a non-trivial vortex configuration in co-dimension three.
The tachyon potential is assumed to be such that the minima of $T$ have 
the eigenvalues $(T_0,-T_0)$, so that the vacuum manifold is 
${\cal V} = {\rm U}(2)/({\rm U}(1)\times {\rm U}(1))=S^2$. The possible
stable vortex configurations  $T({\bf x})$, depending on $3$ coordinates 
$x^i$ transverse to the ($p-2$)-dimensional core of the vortex, are classified by the
non-trivial embeddings of the ``sphere at infinity'' $S^{2}_\infty$ into the
vacuum manifold, namely by $\pi_{2}({\cal V}) = {\bf Z}$.
\par
Apart from  the ``center of mass'' ${\rm U}(1)$ 
subgroup, we are in the situation of the Georgi-Glashow model,
where the tachyon field $T({\bf x})= T^a({\bf x}) \sigma^a$ 
($\sigma^a$ being the Pauli matrices) sits in the adjoint of  ${\rm SU}(2)$, 
and the vacuum manifold is described by $T^a T^a = T_0^2$.
The vortex configuration of winding number one of the tachyon, which is the 't Hooft-Polyakov
monopole, is accompanied by a non-trivial SU(2) gauge field:
\begin{eqnarray}
\label{thmon}
T({\bf x}) & = & f(r) \sigma_a x^a~,\nonumber\\
{\cal A}^a_i({\bf x}) & = & h(r) \epsilon^a{}_{ij} x^j~,
\end{eqnarray}
where $r$ is the radial distance in the three transverse directions; 
the prefactors $f(r)$ and $h(r)$ go to constants for $r\to 0$, while
$f(r)\sim T_0/r$ and $h(r)\sim 1/r^2$ for $r\to\infty$.
The field-strength in the unbroken ${\rm U}(1)$
direction,
\begin{equation}
\label{m5}
{\cal G}_{ij} = {T^a\over T_0} {\cal F}^a_{ij}~~,
\end{equation}
corresponds to a
non-trivial U(1) bundle on the sphere at infinity, {\it i.e.} the magnetic 
charge $g = \int_{S^{2}_\infty} {\cal G}$ is non-zero (and in fact equals 
the winding number of the vortex in appropriate units).
The magnetic charge density 
in the transverse directions, defined by $d{\cal G} = \rho({\bf x}) d^3x$, 
is concentrated at the core of the vortex 
solution. In the zero size limit,  there is a point-like magnetic charge
at the location of the core: $\rho({\bf x}) = g\, \delta^3({\bf x} - {\bf x}^0)$.
\par
The  WZ action (\ref{wznonbps}) for the D$p$-brane can be rewritten as
\begin{equation}
\label{m6}
a\int_{p+1} C\wedge d\,{\rm Tr}\{ T\, {\rm e}^{2\pi\alpha' {\cal F}}\}
\wedge{\rm e}^{2\pi\alpha' {\hat F} + B} \wedge 
\sqrt{\hat{A}(R_T)/\hat{A}(R_N)}~,
\end{equation}
where we have split the U(2) field-strength into its SU(2) part ${\cal F}$
and its U(1) part $\hat F$. Inserting the 't Hooft-Polyakov configuration
for the tachyon and the SU(2) gauge field, we see that Eq.~(\ref{m6}) involves 
the curl $d{\cal G}$ of the magnetic monopole field 
${\cal G}= T^a {\cal F}^a$; we end up with
\begin{equation}
\label{m7}
2\pi\alpha' a T_0 \int_{p+1}  C\wedge \rho({\bf x}) d^3x\wedge
{\rm e}^{2\pi\alpha' {\hat F} + B} \wedge 
\sqrt{\hat{A}(R_T)/\hat{A}(R_N)}~.
\end{equation}
We have a distribution of D($p-3$)-brane charge localized at the 
core of the vortex; in particular, in the limit of zero-size core we recover
the R-R couplings (\ref{wznonbps}) of a BPS D($p-3$)-brane that supports the 
U(1) gauge field $\hat F$.
\par
Since the minimal magnetic charge $g$ is $4\pi$ in our units, Eqs.~(\ref{m2})
and (\ref{m7}) lead to the expected ratio $4\pi^2\alpha'$ for the R-R charges of 
D($p-3$)-~and D($p-1$)-branes.
\par
The mechanism described above generalizes
to the reduction of a non-BPS D$p$-brane to a D$(p-2k-1)$-brane via
tachyon condensation, described in \cite{horava}. In this case, it is 
convenient to start with $2^k$ unstable D$p$-branes. The 
tachyon field, which sits in the adjoint of U$(2^k)$, can be in a vortex
configuration, accompanied by non-trivial gauge fields, such that
there is a non-zero generalized magnetic charge 
$\int_{S^{2k}_\infty}{\rm Tr}\{T ({\cal F})^k\}$. The WZ action (\ref{wznonbps})
then contains the generalized magnetic charge density
$d\,{\rm Tr}\{T {\cal F}^k\}= \rho({\bf x}) d^{2k+1}x$, localized at the
core of the vortex. In the zero-size limit, we
are left with the WZ action for a D$(p-2k-1)$-brane.
\par
The form of the R-R couplings (\ref{wznonbps}) can be checked 
directly by computing the 
corresponding string scattering amplitudes \cite{ournonbps}.
\begin{acknowledgments}
B.C. and F.R. would like to thank the organisers for a very nice school, and for
financial support. This work was supported by the European Commission TMR 
programme ERBFMRX-CT96-0045. B.C. is Aspirant FWO-Vlaanderen.
\end{acknowledgments}
\end{document}